\documentclass{elsart}
\usepackage{epsfig}
\newcommand{\be}{\begin{eqnarray}}
\newcommand{\ee}{\end{eqnarray}}

\newcommand{\real}{\mbox{{\rm I\hspace{-2truemm} R}}}

\begin{document}
\begin{frontmatter}
\title{Black Hole Evaporation and Large Extra Dimensions}
\author[a1]{Roberto Casadio}
\author[a2]{and Benjamin Harms}
\address[a1]{Dipartimento di Fisica, Universit\`a di Bologna,
and I.N.F.N., Sezione di Bologna,
via Irnerio 46, I-40126 Bologna, Italy
\thanksref{e1}}
\address[a2]{Department of Physics and Astronomy,
The University of Alabama,
Box 870324, Tuscaloosa, AL 35487-0324, USA
\thanksref{e2}}
\thanks[e1]{Email: casadio@bo.infn.it}
\thanks[e2]{Email: bharms@bama.ua.edu}

\begin{abstract}
We study the evaporation of black holes in space-times with
extra dimensions of size $L$.
We first obtain a potential which describes the
expected behaviors of very large and very small black holes and
then show that a (first order) phase transition, possibly signaled by
an outburst of energy, occurs in the system when the horizon shrinks
below $L$ from a larger value.
This is related to both a change in the topology of the horizon and
the restoring of translational symmetry along the extra dimensions.
\end{abstract}
\end{frontmatter}
Since Hawking's semiclassical computation \cite{hawking}, one of the
most elusive riddles of contemporary theoretical physics has been to
understand black hole evaporation in a fully dynamical framework which
accounts for the backreaction of the emitted particles on the space-time
geometry.
Intrinsically related to this (technically and conceptually) difficult
issue is the role played by quantum gravity, since, by extrapolating from
the semiclassical picture, one expects that black holes are capable of
emitting particles up to any physical mass scale, including the Planck
mass $m_p=\ell_p^{-1}$.
Therefore, black holes appear as the most natural window to look through
for any theory involving quantum gravity.
\par
In a series of papers \cite{r1,mfd} the point of view of
statistical mechanics \cite{st} was taken and the analysis based on
the well known fact that the canonical ensemble cannot be consistently
defined for a black hole and its Hawking radiation (conventionally
viewed as point-like).
Instead, the well posed microcanonical description led to
the conclusion that black holes are (excitations of) extended objects
($p$-branes), a gas of which satisfies the bootstrap condition.
This yielded the picture in which a black hole and the emitted
particles are of the same nature and an improved law of black hole
decay which is consistent with unitarity (energy conservation).
The weakness of the statistical mechanical analysis is that it does
not convey the geometry of the space-time where the evaporation takes
place: although the luminosity of the black hole can be computed as a
function of its mass, the backreaction on the metric remains an
intractable problem \cite{mfd}.
\par
This scenario has received support from investigations in fundamental
string theory, where it is now accepted that extended D$p$-branes are
a basic ingredient \cite{polchinski}.
States of such objects were constructed which represent black holes
\cite{vafa} and corroborate \cite{maldacena} the old idea that the
area of the horizon is a measure of the quantum degeneracy of the black
hole \cite{st}.
However, this latter approach works mostly for very tiny black holes
and suffers from the same shortcoming that the determination of the
space-time geometry during the evaporation is missing.
\par
It appears natural, in the framework of string theory, to consider
the case when the black hole is embedded in a space-time of higher
dimensionality.
Indeed, the interest in models with extra spatial dimensions has been
recently revived since they deliver a possible solution to the hierarchy
problem without appealing to supersymmetry (see \cite{arkani} and
references therein).
One qualitatively views the four dimensional space-time as a D3-brane
embedded in a bulk space-time of dimension $4+d$.
Since matter is described by open strings with endpoints on the
D$p$-branes, one expects that matter fields are confined to live on the
D3-brane in the low energy limit ({\em e.g.}, for energy smaller than
the electroweak scale $\Lambda_{EW}$), while gravity, being mediated by
closed strings, can propagate also in the bulk.
The $d$ extra spatial dimensions can be either compact \cite{arkani}
or infinitely extended \cite{RS}.
Black holes in the former scenario were considered in
\cite{argyres} and some light on the non-compact case was
shed in \cite{chamblin} (see also \cite{emparan} and
References therein).
\par
For compact extra dimensions of typical size $L$, states of
the gravitational field living in the bulk have $d$ momentum
components quantized in units of $2\pi/L$.
For an observer living on the D3-brane, they then appear as
particles with mass
\be
m^{(\vec n)}
\sim m_p\,{\ell_p\over L}\,\sum_{i=1}^d n_i
\ ,
\ee
where all $n_i=0$ for the (ground state) massless excitations
corresponding to four dimensional gravitons, and states with
some $n_i>0$ are known as Kaluza-Klein (KK) modes which induce
short-range deviations from Newton's law.
The masses $m^{(\vec n)}$ can be relatively small and the energy of
matter confined on the D3-brane is prevented from leaking into
the extra dimensions by the small coupling ($\sim m_p^{-1}$)
between KK states of gravity and the matter energy-momentum
tensor.
This implies that processes involving particles within the
standard model should result in energy loss into the extra
dimensions and other phenomena potentially observable only
above the TeV scale \cite{arkani}.
However, this protection is not effective with Hawking
radiation, since a black hole can evaporate into all
existing particles whose masses are lower than its temperature,
thus providing an independent way of testing the existence of
extra dimensions.
\par
The easiest way to estimate deviations from Newton's law is to
evaluate the potential generated by a point-like source of (bare)
mass $M$ by means of Gauss' law \cite{arkani}.
Let us denote by $r_b$ the usual area coordinate on the four
dimensional D3-brane.
For distances $r_b\gg L$ one then recovers the standard form
\be
V_{(4)}=-G_N\,{M\over r_b}
\ ,
\label{V>}
\ee
where $G_N=m_p^{-2}$ is Newton's constant in four dimensions.
For $r_b<L$ one has
\be
V_{(4+d)}=-G_{(4+d)}\,{M\over r_b^{1+d}}
\ ,
\label{V<}
\ee
with $G_{(4+d)}=M_{(4+d)}^{-2-d}=L^d\,G_N$.
This implies that the huge Planck mass
$m_p^2=M_{(4+d)}^{2+d}\,L^d$ and, for sufficiently large
$L$ and $d$, the bulk mass scale $M_{(4+d)}$ (eventually
identified with the fundamental string scale) can be as small
as $1\,$TeV.
Since
\be
L\sim \left[{1\,{\rm TeV}/M_{(4+d)}}\right]^{1+{2\over d}}\,
10^{{31\over d}-16}\,{\rm mm}
\ ,
\label{tev}
\ee
requiring that Newton's law not be violated for distances larger
than $1\,$mm restricts $d\ge 2$ \cite{arkani}.
\par
A form for the potential which yields the expected behaviour at
both small and large distance and ensures that a test particle
on the D3-brane does not experience any discontinuity in the
force when crossing $r_b\sim L$ \footnote{An important effect to
be tested by the next generation of table-top experiments, see,
{\em e.g.}, \cite{long}.} is given by
\be
V=-G_N\,{M\over r_b}\,\left[1+\sum_{n=1}^d\,C_n\,
\left({L\over r_b}\right)^n\right]
\ ,
\label{V}
\ee
where $C_n$ are numerical coefficients (possibly functions of $M$,
see below).
The second (Yukawa) contribution in (\ref{V}) can be
related to the exchange of (massive) KK modes and, hence, encodes
tidal effects from the bulk due to the presence of the mass $M$
on the D3-brane.
We further note that the case $d=2$ can be used to make estimates
for one non-compact extra dimension \cite{emparan}, provided one
introduces an effective size $L^2\sim -M_{(5)}^3/\Lambda$, with
$\Lambda$ the (negative) cosmological constant in the bulk
AdS$_5$ \cite{RS}.
\par
The behavior of both very large ($R_H\gg L$) and very small
($R_H\ll L$) black holes is by now relatively well understood.
In the former case, one can unwrap the compact extra dimensions and
regard the real singularity as spread along a (black) $d$-brane
\cite{horowitz} of uniform density $M/L^d$, thus obtaining the
Schwarzschild metric on the orthogonal D3-brane, in agreement with
the weak field limit $V_{(4)}$, and an approximate ``cylindrical''
horizon topology $S^2\times \real^d$.
In the latter case a solution is known \cite{chamblin}, for one
infinite extra dimension \cite{RS}, which still has the form of a
black string extending all the way through the bulk AdS$_5$.
However, this solution is unstable \cite{gregory} and believed to
further collapse into one point-like singularity \cite{chamblin}.
This can be also argued from the observation that the Euclidean
action of a black hole is proportional to its horizon area and is
thus minimized by the spherical topology $S^{2+d}$.
Hence, small black holes are expected to correspond to a
generalization of the Schwarzschild metric to $4+d$ dimensions
\cite{myers} and should be colder and (possibly much) longer
lived \cite{argyres}.
\par
However, there is still a point to be clarified for small black
holes, namely one should find an explicit matching between the
spherical metric (for $r_b<L$) and the cylindrical metric
(for $r_b>L$).
This is not a trivial detail, since the ADM mass of a spherical
$4+d$ dimensional black hole is zero as seen from the D3-brane
because there is no $1/r_b$ term in the large $r_b$ expansion of
the time-time component of the metric tensor \cite{myers}.
Thus, one concludes that the 4 dimensional ADM mass of a small
black hole can be determined as a function of the $4+d$
dimensional mass parameter only after such a matching is provided
explicitly.
Even less is known about black holes of size $R_H\sim L$, and a
complete description is likely to be achieved only by solving the
entire set of field equations for an evaporating black hole
in $4+d$ dimensions.
\par
Instead of tackling this intractable backreaction problem, we
extrapolate from the weak field limit on the D3-brane given in
(\ref{V}) the (time and radial components of the) metric in
$4+d$ dimensions as
\be
g_{tt}&\simeq& -1-2\,V(r)
\nonumber \\
&=&
1-2\,G_N\,{M\over r}\,\left[1+\sum_{n=1}^d\,C_n\,
\left({L\over r}\right)^n\right]
\nonumber \\
&&
\label{met}
\\
g_{rr}&\simeq&-g_{tt}^{-1}
\ ,
\nonumber
\ee
where $r$ now stands for the area coordinate in $4+d$
dimensions.
This yields $M$ as the ADM mass of the black hole (see
\cite{maartens} for a similar solution in the scenario
of \cite{RS}) and the radius of the horizon is determined by
\be
g_{tt}=0\ \Rightarrow\
R_H=2\,G_N\,M\,\left[1+\sum_{n=1}^d\,
C_n\,\left({L\over R_H}\right)^n\right]
\ .
\label{R_H}
\ee
The above {\em ansatz\/} does not provide an exact solution of
vacuum Einstein equations, since some of the components
of the corresponding Einstein tensor in $4+d$ dimensions
$G_{ij}=8\,\pi\,G_{(4+d)}\,T_{ij}\not =0$.
However, it is possible to choose the coefficients $C_n$ (as
functions of $M$) in such a way that the ``effective matter
contribution'' $T_{ij}$  from the region outside the black hole
horizon is small.
In particular, one can require that the contribution to the ADM
mass be negligible,
\be
m\equiv\int_{R_H}^\infty d^{4+d}x\,
T^t_{\ t}=
{1\over 8\,\pi\,G_{(4+d)}}\,
\int_{R_H}^\infty d^{4+d}x\,G^t_{\ t}(\{C_n\})
\ll M
\ .
\ee
In this sense one can render the above metric a good approximation
to a true black hole in $4+d$ dimensions.
One should also recall that the black hole must be a classical
object, to wit its Compton wavelength
$\ell_M\sim \ell_p\,(m_p/M)\ll R_H$.
Further, once Hawking radiation is included, its backreaction
on the metric at small $r$ is likely to be significant for
$R_H\ll L$, so that a true vacuum solution would not be
practically much more useful.
\par
When $R_H\gg L$ one has $r\simeq r_b$, therefore, the metric
(\ref{met}) is approximately cylindrically symmetric (along the
extra dimensions).
Further, on setting all $C_n=0$ yields $m=0$, and
Eq.~(\ref{R_H}) coincides with the usual four dimensional
Schwarzschild radius $R_H\simeq 2\,\ell_p\,(M/m_p)$.
Correspondingly one has the inverse Hawking temperature
($\beta_H=T_H^{-1}$) and Euclidean action \cite{hawking,st}
\be
\beta_H^>&\simeq& 8\,\pi\,\ell_p\,(M/m_p)
\label{B>}
\\
S_E^>&\simeq&4\,\pi\,(M/m_p)^2\simeq{A}_{(4)}/4\,\ell_p^2
\ ,
\label{SE>}
\ee
where ${A}_{(D)}$ is the area of the horizon in $D$ space-time
dimensions and the condition $R_H\gg L$ translates into
\be
M\gg m_p\,(L/\ell_p)\equiv M_c
\ ,
\label{Mc}
\ee
({\em e.g.}, $M_c\sim 10^{27}\,$g for $L\sim 1\,$mm).
The fact that the extra dimensions do not play any significant role at
this stage is further confirmed by $T_H^>\ll 2\,\pi/L\equiv m^{(1)}$
(the mass of the lightest KK mode), therefore no KK particles can
be produced.
\par
For $R_H\ll L$, one can again take advantage of the coefficients
$C_n$ to lower $m$ as much as possible.
For instance, for $d=2$ and $M=10^{15}\,$g ($\ll M_c$) the values
$C_2=-C_1=1$ yield $m\sim 10^{-3}\,M$
(more details will be given in \cite{forth}).
Eq.~(\ref{R_H}) then leads to
\be
R_H\simeq \left(2\,C_d\,L^d\,G_N\,M\right)^{1\over 1+d}
\ ,
\label{R_H<}
\ee
and the consistency conditions $\ell_M\ll R_H\ll L$ hold for
\be
m_p\,(\ell_p/L)^{d\over 2+d}\ll M\ll M_c
\ .
\ee
Since we have assumed that the spherical symmetry extends to
$4+d$ dimensions, one obtains \cite{argyres}
\be
&&\beta_H^<\sim L\,(M/M_c)^{1\over 1+d}
\\
&&S_E^<\sim\left(L/\ell_p\right)^2\,\left(M/M_c\right)^{2+d\over 1+d}
\sim{A}_{(4+d)}/\ell_p^2\,L^d
\ ,
\label{SE<}
\ee
which reduce back to (\ref{B>}) and (\ref{SE>}) if one pushes
down $L\to \ell_p$ ($M_c\to m_p$).
For $M\ll M_c$, the temperature $T_H^<$ is sufficient to excite KK
modes, although it is lower than that of a four dimensional black
hole of equal mass.
Correspondingly, the Euclidean action $S_E^<(M)\ge S_E^>(M)$,
yielding a smaller probability \cite{hawking,r1}
\be
P\sim\exp\left(-S_E\right)
\ .
\label{P}
\ee
for the Hawking particles ``to come into existence'' in the $4+d$
dimensional scenario.
\par
The luminosity of a black hole (provided microcanonical corrections
are negligible \cite{mfd,forth}) can be approximated by employing the
canonical expression \cite{hawking}
\be
F_{(D)}\simeq{A}_{(D)}\,
\sum_s\,\int_0^\infty {\Gamma_s\,d\omega^D
\over e^{\beta_H\,\omega}\mp 1}
={A}_{(D)}\,N_{(D)}\,T_H^{\,D}
\ ,
\label{L}
\ee
with $\Gamma$ the grey-body factor and $N$ a coefficient which
depends upon the number of available particle species $s$ with mass
smaller than $T_H$.
For small black holes $D=4+d$ and $T_H=T_H^<$, hence
$F_{(4+d)}\sim (M/M_c)^{-{2\over 1+d}}$ is far less
intense than $F_{(4)}\sim (M/M_c)^{-2}$ obtained from $T_H^>$
\cite{argyres}.
It also follows from (\ref{L}) that $F_{(D)}$ as a function of
$R_H$ (not $M$) depends on $D$ only through $N_{(D)}$.
Therefore, the energy emitted into KK modes must be a small
fraction of the total luminosity,
\be
{F_{KK}\over F_{(4+d)}}\sim
{N_{KK}\over N_{(4)}+N_{KK}}\ll 1
\ ,
\ee
because the number of degrees of freedom of KK gravitons
($\sim N_{KK}$) is much
smaller than the number of particles in the standard
model ($\sim N_{(4)}$), and energy conservation in $4+d$ dimensions
requires that $F_{(4+d)}=F_{(4)}+F_{KK}$
(see also \cite{emparan} for similar arguments).
However, one might question the efficiency of the confining
mechanism for matter on the D3-brane at very high
energy\footnote{There is indeed evidence that
only the zero modes of standard model fields can be confined,
thus allowing matter to leak in the bulk (see, {\em e.g.},
\cite{alex}).}
(say greater than $\Lambda_{EW}$).
One should then include bulk standard model fields among KK
modes and consider $N_{KK}$ as a growing function of the
temperature.
If this is the case, the luminosity will greatly increase and
the ratio ${F_{KK}/F_{(4+d)}}$ will eventually approach unity
when $T_H^<>\Lambda_{EW}$.
\par
An inspection of the Klein-Gordon equation in $4+d$ dimensions
supports the conclusion that the luminosity of small black holes
is relatively fainter.
In fact, there is a potential barrier in the equation governing the
radial propagation outside the horizon of the form \cite{forth}
\be
W\sim
{d\over 2}\,\left({d\over 2}+1\right)\,{1\over r^2}\,
\left[1-\left({R_H\over r}\right)^{1+d}\right]^2
\ ,
\ee
which all Hawking particles have to tunnel through in order to
escape.
One can roughly reproduce $W$ by assuming that all particles are
emitted with an effective angular momentum $\sim l+d/3$,
which substantially lowers the grey-body factor \cite{page}.
\par
Let us now consider the case when $R_H\sim L$
({\em i.e.}, $M\sim M_c$) and try to
understand what happens as the horizon shrinks below the size of
the extra dimensions.
It is easy to solve (\ref{R_H}) for the ``toy model'' $d=1$ and
estimate the relation between the ADM mass and the radius of the
horizon at all scales
\be
R_H=G_N\,M\,\left(1+\sqrt{1+(2\,L/G_N\,M)}\right)
\ .
\ee
This leaves only the topology of the horizon to be specified
for $2\,R_H\le L$.
According to the area law \cite{st}
$S_E\sim {A}_{(4+d)}/16\,\pi\,G_{(4+d)}$,
one has for the ``cylinder'' $S^2\times \real$ and
for the three sphere $S^3$ respectively
\be
S_E^c\sim R_H^2/8\,\ell_p^2
\ ,
\ \ \ \
S_E^s\sim R_H^3/16\,\ell_p^2\,L
\ ,
\label{SE1}
\ee
and the ratio $S_E^c/S_E^s\sim 4$ for $2\,R_H=L$,
so that the spherical topology is favored once the horizon
has become small enough to ``close up'' along the extra dimension.
The inverse temperature in the two cases is given by
\be
\beta_H^c\sim {R_H\over 4\,\ell_p^2}\,{\partial R_H\over\partial M}
\ ,\ \ \ \
\beta_H^s\sim
{3\,R_H^2\over 16\,\ell_p^2\,L}\,{\partial R_H\over\partial M}
\ ,
\ee
and is discontinuous as well, since $\beta_H^c/\beta_H^s\sim 2/3$
for $2\,R_H=L$.
The physically interesting case $d=2$ is algebraically more involved
but the qualitative picture is the same as for $d=1$ if the
topology of the extra dimensions is $\real^2$ \cite{forth}.
\par
We can now get some further physical insight by appealing to statistical
mechanics.
Although the use of the canonical ensemble is known to be incorrect,
we utilize it in the analysis below in order to conform to
the standard description and introduce a partition function for the
black hole and its Hawking radiation as the Laplace transform of the
microcanonical density of states \cite{huang,r1,mfd},
\be
Z_\beta\sim\int dM\,e^{-\beta\,M}\,e^{S_E(M)}
\ ,
\label{Z}
\ee
where we have estimated the internal degeneracy of a black hole as the
inverse of the probability (\ref{P}), in agreement with the area
law \cite{st,maldacena} and the bootstrap relation \cite{r1,mfd}.
Since the entropy $S=\beta^2\,(\partial F/\partial\beta)$, where
$\beta\,F=-\ln Z$ is the (Helmholtz) free energy, it follows from
(\ref{SE1}) that there is a discontinuity in the first derivative
of $F$ at
\be
\beta_c\simeq \beta_H^s(M_c)\sim L\sim 1/m^{(1)}
\ .
\ee
This behavior is characteristic of a first order phase transition
\cite{huang}:
In the cold phase, $T_H<\beta_c^{-1}$, the system appears
``condensed'' into the low energy standard model fields living
on the four dimensional D3-brane and translational invariance in
the $d$ extra directions is therefore broken by the D3-brane
itself.
For $T_H>\beta_c^{-1}$ translational invariance begins to restore,
and the system starts spreading over all bulk space, with D3-brane
vibrations acting as Nambu-Goldstone bosons which give mass to the
KK modes\footnote{One might argue that it is the cylindrical topology
which is translationally symmetric along the extra dimensions,
however for $T_H<T_c$ there are no physical particles in the bulk,
and such a symmetry remains a purely geometrical concept (as well
as the bulk itself) with no dynamical counterpart.}
\cite{arkani}.
\par
\begin{figure}
\centerline{\epsfxsize=300pt
\epsfbox{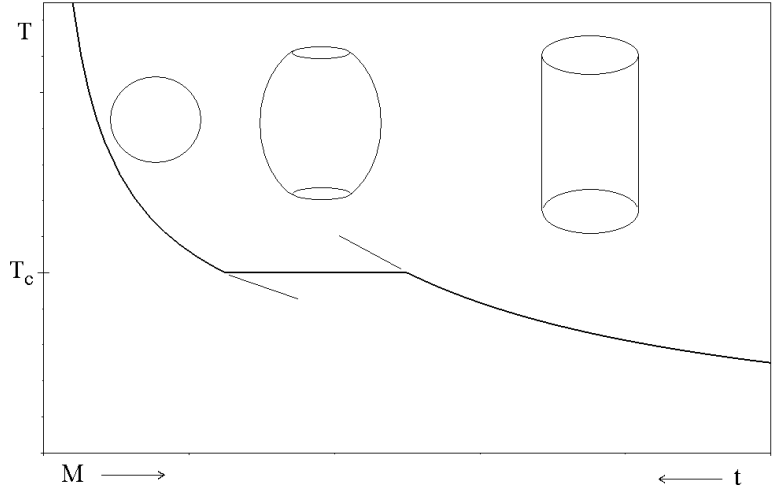}}
\caption{Sketch of the equation of state $T=T(M)$ and corresponding
topology of the horizon.
Mass increases to the right (time lapses towards the left).}
\label{fig}
\end{figure}
Of course, in the statistical mechanical analysis the time is missing,
and one can more realistically think that, during the evaporation of
an initially large black hole, the horizon starts to bend in
the extra dimensions when $T_H^c$ approaches $T_H^s(M_c)$ (from below).
If the temperature remains constant (and approximately equal to the
lower value $T_H^s(M_c)\equiv T_c$) during the transition, then
a qualitative picture of the effect is given in Fig.~\ref{fig}.
The change in the topology of the horizon could then be accompanied by
a burst of energy to get rid quickly of the excess mass
(equal to the width of the plateau in Fig.~\ref{fig}).
\par
In the above approximation the specific heats in the cold and hot phases
are respectively given by
\be
C_V^>\sim -\left({M/m_p}\right)^2
\ ,\ \ \
C_V^<\sim -L\,M\,\left({M/M_c}\right)^{1\over 1+d}
\ ,
\ee
Since $C_V^<$ is negative, one is eventually forced to employ the
microcanonical description when the temperature becomes
too hot\footnote{In order to make sense of the partition function,
one should introduce an ultra-violet cut off $\Lambda_{UV}$ to render
the (divergent) integral (\ref{Z}) finite.
Then, the canonical description is a good approximation only for
$T\ll\Lambda_{UV}$ \cite{forth}},
although the fact that $|C_V^<|<|C_V^>|$
suggests that the extra dimensions make the inconsistency of the
canonical description milder \cite{forth}.
%
%
\par
This work was supported in part by the U.S. Department of Energy under
Grant no.~DE-FG02-96ER40967 and by NATO grant no.~CRG.973052.
\end{document}